\documentclass[12pt]{iopart}
\usepackage{amsfonts}   % note how statements can be commented out
\usepackage{amssymb}
\usepackage{graphicx,epstopdf}   % for figures
\newtheorem{theorem}{Theorem}
\usepackage{iopams}
\usepackage{setspace}

\begin{document}

\title{Inferring  the smoothness of  the autocorrelation function from that of the initial state}

\author{Yang K L$^{1, 2}$ and Zhang J M$^{1, 2} $}

\address{$^1$ Fujian Provincial Key Laboratory of Quantum Manipulation and New Energy Materials,
College of Physics and Energy, Fujian Normal University, Fuzhou 350007, China}
\address{$^2$ Fujian Provincial Collaborative Innovation Center for Optoelectronic Semiconductors and Efficient Devices, Xiamen 361005, China}
% \ead{wdlang06@163.com}
\vspace{10pt}
\begin{indented}
\item[]June 2019
\end{indented}

\begin{abstract}
We point out that by the ``smoothness means fast decay'' principle in Fourier analysis, it is possible to infer the smoothness (or nonsmoothness) of the autocorrelation function from a mere glimpse of the initial state. Specifically, for a generic system with smooth eigenstates, the smoother an initial state is, the faster its decomposition coefficients with respect to the eigenstates of the system decay, and in turn the smoother the autocorrelation function is. The idea is illustrated with three increasingly smooth functions in an infinite square well. By using the Mellin transform, we also find that the nonsmoothness or singularity of the initial state affects the short-time behavior of its autocorrelation function. In particular, a sufficiently nonsmooth or singular initial state could decay in a nonquadratic power law,  with the exponent continuously tunable. We now understand the periodic cusps of the autocorrelation function in the quench dynamics of a Bloch state, which was observed previously [Zhang and Yang, EPL 114, 60001 (2016)].
\end{abstract}
\vspace{2pc}

%\maketitle

\section{Introduction}

In quantum dynamics, a fundamental quantity is the so-called autocorrelation function (or survival amplitude) \cite{robinett1, nauenberg}, which measures the recurrence (or survival) of an initial state. Let the Hamiltonian of the system be $H$, and the initial state be $|\psi_i\rangle $. The autocorrelation function associated with this particular state is defined as the overlap between it and its time-evolved state  $|\psi(t) \rangle = e^{-i H t }|\psi_i \rangle  $,
\begin{eqnarray}\label{defa}
  A(t) = \langle \psi_i  | \psi (t) \rangle= \langle \psi_i |e^{-i H t} | \psi_i \rangle .
\end{eqnarray}
So, $|A|^2$ is the survival probability of the initial state in the time-evolved state. This makes it the quantity of primary interest in quantum decay.
Introducing the eigenstates $\{ |\phi_n \rangle \}$ and the corresponding eigenenergies $\{E_n\}$ of $H$, we have
 \begin{eqnarray}\label{forma}
  A(t)  &=& \sum_{n=1}^\infty  |c_n|^2 e^{-i E_n t } ,
\end{eqnarray}
where the coefficients $c_n = \langle \phi_n | \psi_i \rangle $ come from the decomposition
\begin{eqnarray}\label{decomp}
  |\psi_i\rangle &=& \sum_{n=1}^\infty c_n |\phi_n \rangle  .
\end{eqnarray}
From the point of view of Fourier analysis, here two procedures working in opposite directions are involved. First, in the forward direction, as in (\ref{decomp}),  the initial state is decomposed with respect to the orthonormal basis $\{|\phi_n \rangle \}$, and the coefficients $c_n $ are obtained; then, in the backward direction, as in (\ref{forma}), the autocorrelation function is constructed as a Fourier series with $|c_n|^2 $ being the coefficients.

Now, we note that in Fourier analysis, a very general and very important principle is that the smoothness of a function is linked to the decay rate of its Fourier coefficients---the smoother the function is, the faster the Fourier coefficients decay \cite{stein}. While this principle can only be appropriately understood  in terms of rigorous theorems (see below for some), it is in agreement with the intuition that to fit a jump or a cusp we need fast oscillating terms.

By this ``smoothness means fast decay'' principle (and its reverse), an initial state sufficiently smooth will yield a series $\{c_n \}$ decaying sufficiently fast, which in turn will lead to a sufficiently smooth autocorrelation function in time. On the contrary, an initial state with jumps or cusps will yield a series decaying relatively slow, and consequently will result in an autocorrelation function less smooth.

It is the purpose of this paper to verify this point. In the following section, we take the infinite square well potential as our setting to investigate the connection between the smoothness of an initial state and that of its autocorrelation function. We shall examine three increasingly smooth functions, which have discontinuity either at the zeroth, or the first, or the second derivatives. We shall see that the corresponding autocorrelation functions are increasingly smooth. Interestingly, for the wave function discontinuous at the zeroth order, its autocorrelation function realizes the famous \emph{Riemann function}, which is continuous everywhere but differentiable only at countably many points.

Incidentally, in the study, we noticed also that the three autocorrelation functions differ not only in their differentiability, but also in their short-time behaviors. Visually apparent is that, the autocorrelation function corresponding to the  initial state discontinuous at the zeroth order exhibits a cusp at $t=0$, while the other two look much smoother there and seem  more close to being decaying quadratically, as is often assumed in the discussion of quantum Zeno effect. This sharp difference  motivated us to study the asymptotic behaviors of the autocorrelation functions in the limit of  $t\rightarrow 0$. Technically, this is done by using the Mellin transform. The finding is that in (\ref{forma}), the decay mode of the coefficients $\{ |c_n|^2 \}$ is determinant for the short-time asymptotic behavior of the autocorrelation function $A$. In particular, a  series $\{ |c_n|^2 \} $ decaying sufficiently slowly could lead to a fractal power law behavior of the autocorrelation function, which is exactly what is happening behind the cusp aforementioned.

In this tentative work, we of course tend to be illustrative instead of comprehensive, as is evident from the choice of the simplest model in quantum mechanics. Pursuing the theme in more complicated models, especially in higher dimensions, seems both worthy and challenging. We shall discuss these issues in the conclusion part.

\section{Differentiability}
By choosing proper units, we can assume that the well is on the interval $(0, \pi)$, and the Hamiltonian is $H_{{w}} = - \partial^2/\partial x^2 $. The eigenstates are then $  \phi_n  = \sqrt{2/\pi} \sin n x $ and the eigenenergies  are $ E_n = n^2 $ ($n\geq 1$).
For given arbitrary initial state $\psi $, the decomposition $\psi = \sum_n c_n \phi_n $ can be carried out by calculating the coefficients $c_n $ as
\begin{eqnarray}
  c_n &=& \int_0^\pi dx \phi_n (x) \psi(x) .
\end{eqnarray}
But to make use of ready-made results in Fourier analysis, we define the odd function $\tilde{\psi }$ on $(-\pi , \pi )$,
\begin{eqnarray}
  \tilde{\psi}(x) =  \cases{ \psi(x), & $0< x < \pi $, \\ - \psi(-x ), & $-\pi < x < 0 $, \\  }
\end{eqnarray}
and then extend it to the whole axis by periodicity. The reason of introducing $\tilde{\psi }$ is that, if $\psi = \sum_{n\geq 1} c_n \phi_n $, by parity we have automatically  that,
\begin{eqnarray}
  \tilde{\psi } &=& \sum_{n\geq 1} c_n \sqrt{\frac{2}{\pi }} \sin n x = \sum_{n\geq 1} \frac{-i c_n  }{\sqrt{2\pi }} (e^{inx } - e^{-in x }) .
\end{eqnarray}
We then see that the quantity $c_n$ which we are concerned with are proportional to the decomposition coefficients of $\tilde{\psi }$ with respect to the orthonormal basis $ \{ e^{in x }/ \sqrt{2\pi } , n \in \mathbb{Z} \} $ on $(-\pi , \pi)$.

Therefore, the problem reduces to a standard one of the decay rate of the Fourier coefficients of the function $\tilde{\psi}$ in the standard basis $\{ e^{inx}/\sqrt{2\pi } \}$. For the latter problem, several useful theorems are well-known \cite{hardybook}.

\begin{theorem}\label{th1}
If $f$ is $2\pi$-periodic and $f^{(k-1)}$ absolutely continuous on $[0, 2\pi ] $, $f(x) \sim \sum_{n} d_n e^{inx}/\sqrt{2\pi}$, then $d_n = O(1/n^k)$ as $|n| \rightarrow \infty$.
\end{theorem}
Integrating by parts (this is why the condition of absolute continuity is needed), we have
\begin{eqnarray}
  d_n  &=& \frac{1}{\sqrt{2\pi }}\int_{-\pi}^\pi dx f (x) e^{-in x } \nonumber \\
  &=&   \frac{1}{\sqrt{2\pi }} \left [ f(x)\frac{e^{-in x }}{-i n }\bigg |_{-\pi}^\pi +\frac{1}{i n } \int_{-\pi}^\pi dx f'(x) e^{-in x }  \right]\nonumber \\
  &=&  \frac{1}{ i n \sqrt{2\pi }}  \int_{-\pi}^\pi dx f'(x) e^{-in x } .
\end{eqnarray}
Here in the second line, because of the periodicity of the integrand, the boundary contributions cancel. Iterating gives
\begin{eqnarray}\label{dnk}
  d_n  &=&  \frac{1}{ (i n)^k \sqrt{2\pi }}  \int_{-\pi}^\pi dx f^{(k)}(x) e^{-in x } .
\end{eqnarray}
We then have
\begin{eqnarray}
  |d_n| &\leq &  \frac{1}{ n^k \sqrt{2\pi }}  \int_{-\pi}^\pi dx  |f^{(k)}(x) | = O(n^{-k }).
\end{eqnarray}

\begin{theorem}\label{th2}
If $f$ is of bounded variation on $[0, 2\pi ] $, $f(x) \sim \sum_{n} d_n e^{inx}/\sqrt{2\pi} $, then $d_n = O(1/n)$ as $ |n| \rightarrow \infty$.
\end{theorem}
A very short and illuminating proof of this theorem was given by Taibleson in \cite{bounded}.

\begin{theorem}\label{th3}
If $f$ is $2\pi$-periodic, $f^{(k-1)}$ absolutely continuous on $[0, 2\pi ] $, and $f^{(k)}$ is of bounded variation,  $f(x) \sim \sum_{n} d_n e^{inx}/\sqrt{2\pi}$, then $d_n = O(1/n^{k+1})$ as $|n| \rightarrow \infty$.
\end{theorem}
This is just a combination of the two previous theorems. Applying theorem \ref{th2} to the integral in (\ref{dnk}), we get one more factor of $1/n$.
\begin{figure}[tb]
\centering
\includegraphics[width= 0.45\textwidth ]{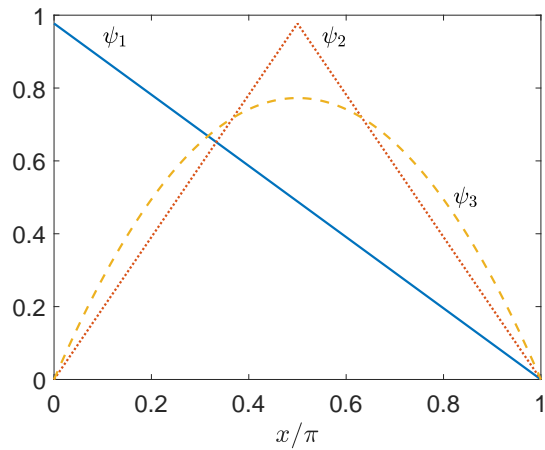}
\caption{(Color online) Graphs of the three initial states in (\ref{threeiniaa})-(\ref{threeinicc}). Note that state $\psi_2$ can be realized as a zero-energy eigenstate if a Dirac delta potential with appropriate strength is placed at $x = \pi/2 $ \cite{zero}. }
\label{fig1}
\end{figure}

With these theorems in hand, we proceed to three concrete states, namely
\begin{eqnarray}
  \psi_1 &=&  \sqrt{\frac{3}{\pi^3}} (\pi - x ) , \label{threeiniaa} \\
  \psi_2 &=& \sqrt{\frac{12}{\pi^3}} \left( \frac{\pi}{2} - |x- \frac{\pi}{2}| \right ) ,\label{threeinibb} \\
  \psi_3 &=&  \sqrt{\frac{30}{\pi^5}} x (\pi - x )  . \label{threeinicc}
\end{eqnarray}
They are illustrated in figure~\ref{fig1}. These states are chosen in the order of increasing smoothness. It is easily seen that the extend function $\tilde{\psi}_1$ has a jump at $x = 0$. The function $\tilde{\psi}_2$ is continuous, but its first derivative has jumps at $x= \pm \frac{\pi}{2}$. As for $\tilde{\psi}_3$, its first derivative is still continuous but its second derivative has a jump at $x= 0 $. All these functions and their derivatives are of bounded variation. Therefore, by theorem \ref{th3}, we expect that their Fourier coefficients $c_n$ are on the order of $O(1/n)$, $O(1/n^2)$, and $O(1/n^3)$, respectively. Indeed, by straightforward calculation, we get
\begin{eqnarray}
  \psi_1 &=&  \sum_{n=1}^\infty \frac{\sqrt{6} }{n \pi} \phi_n  , \label{psiexpanaa} \\
  \psi_2 &=& \sum_{n\geq 1, n\in {odd} }\frac{ 4 \sqrt{6}}{n^2 \pi^2} \sin \frac{n \pi}{2} \phi_n  ,\label{psiexpanbb} \\
  \psi_3 &=&  \sum_{n\geq 1, n\in {odd}}\frac{ 8 \sqrt{15} }{n^3 \pi^3}  \phi_n  . \label{psiexpancc}
\end{eqnarray}
We can then form the autocorrelation functions as
\begin{eqnarray}
  A_1 (t) &=& \frac{6 }{ \pi^2} \sum_{n=1}^\infty \frac{e^{-i n^2 t} }{n^2 }  , \label{autofunsaa}  \\
  A_2 (t) &=&  \frac{ 96}{ \pi^4} \sum_{n\geq 1, n\in {odd}} \frac{e^{-i n^2 t} }{n^4 }    , \label{autofunsbb}  \\
  A_3 (t) &=&  \frac{ 960 }{ \pi^6} \sum_{n\geq 1, n\in {odd}} \frac{  e^{-i n^2 t} }{n^6 }    . \label{autofunscc}
\end{eqnarray}

\begin{figure*}[tb]
\centering
\includegraphics[width= 0.32\textwidth ]{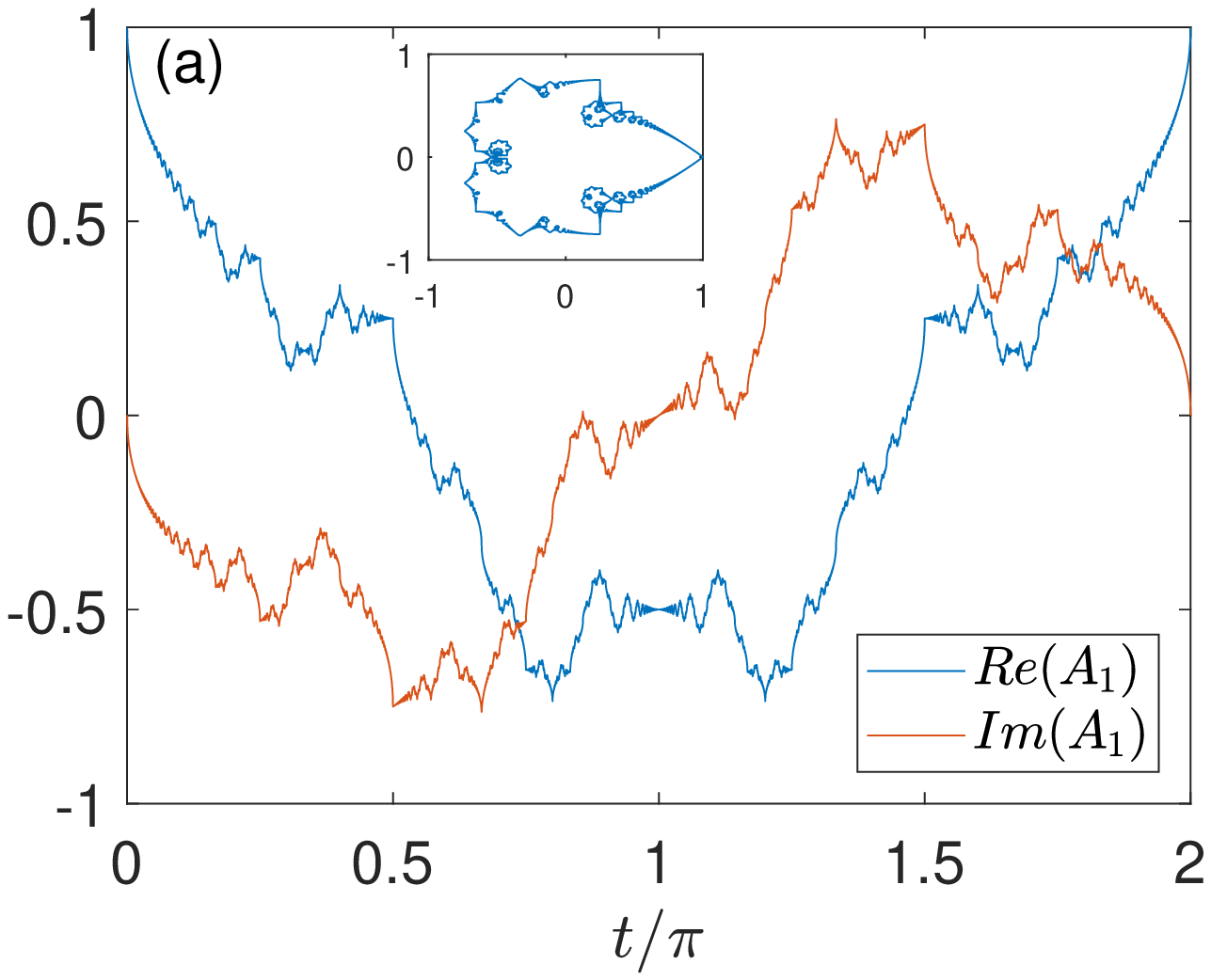}
%\vspace{1mm}
\includegraphics[width= 0.32\textwidth ]{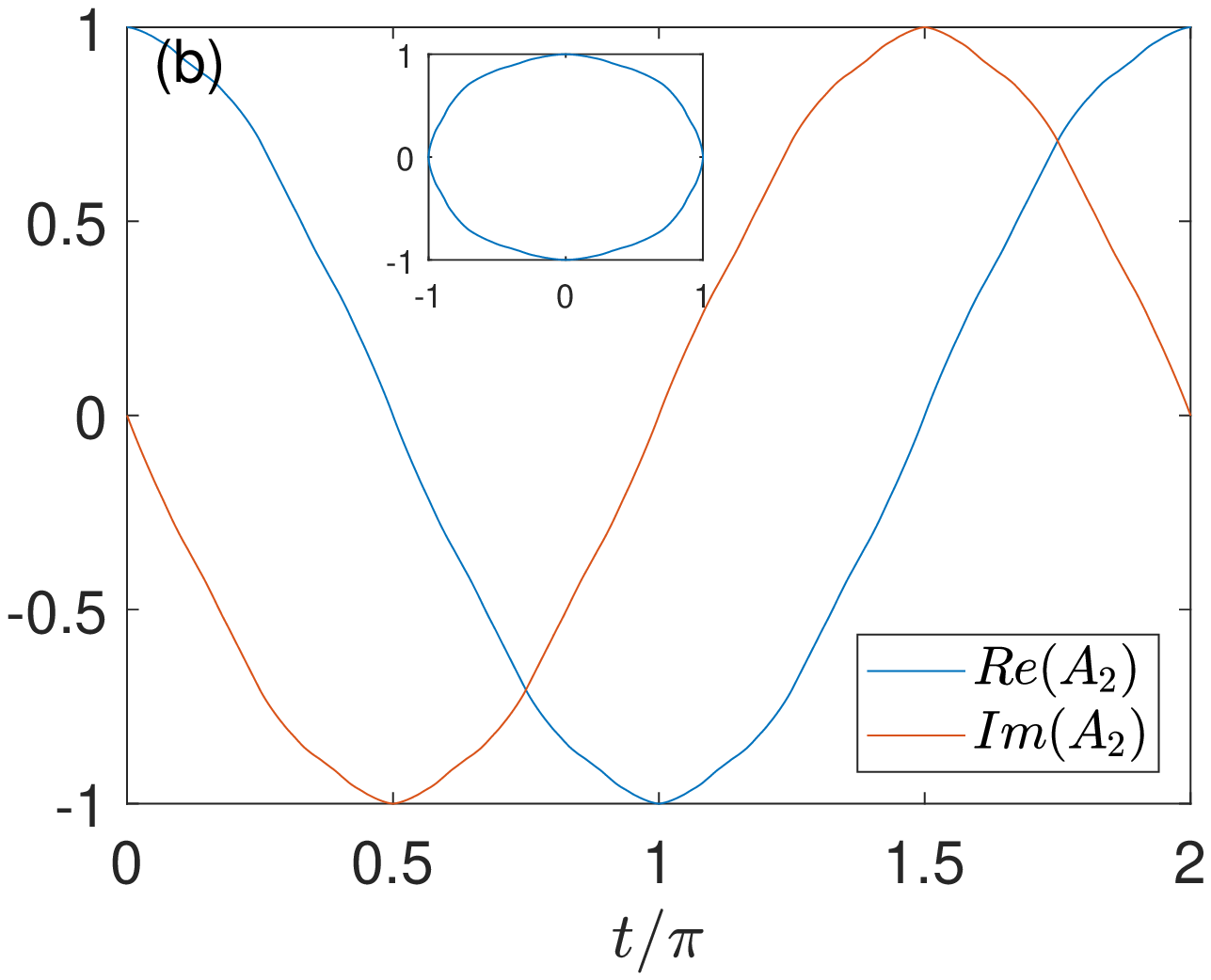}
%\vspace{1mm}
\includegraphics[width= 0.32\textwidth ]{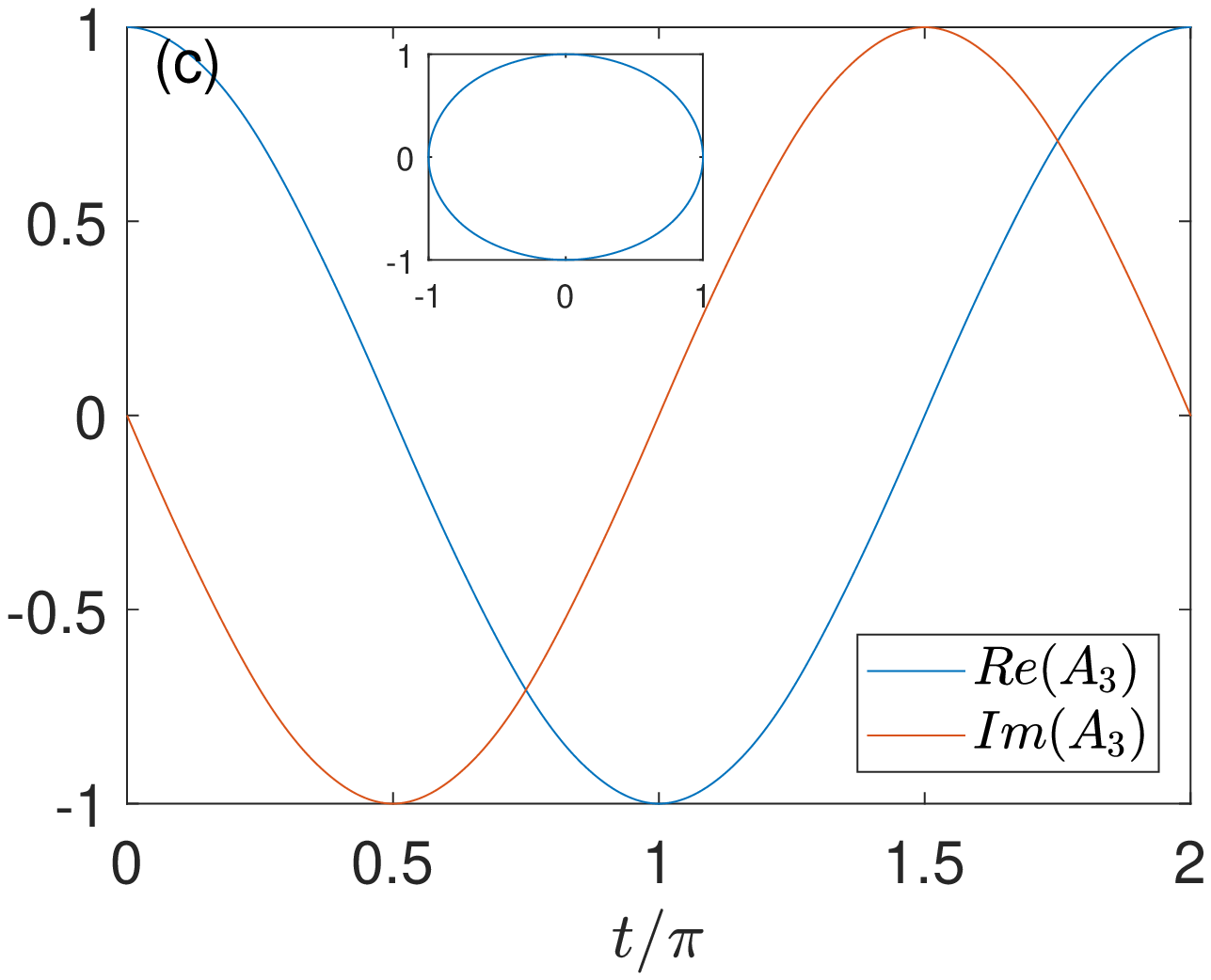}
\caption{(Color online) Real and imaginary parts of the autocorrelation functions $A_i$ ($1\leq i \leq 3 $) in (\ref{autofunsaa})-(\ref{autofunscc}) and (\ref{asymptotics1aa})-(\ref{asymptotics1ff}). As these functions are $2\pi$-periodic, only the parts in the first period are shown. In the insets are the trajectories of $A_i$ on the complex plane. The graphs in (a) are of fractal dimension $5/4$. Note also the $t^{1/2}$-cusps there at $t= 0 $.}
\label{fig2}
\end{figure*}
By Weierstrass' test \cite{knopp}, we see that the three series all converge uniformly on $\mathbb{R}$ and thus the three functions are all continuous. The problem is to what extent they are differentiable. For $A_1$, tentative term-by-term differentiation generates a series converging nowhere, which suggests its poor differentiability. This is supported by its plot in figure~\ref{fig2}(a), where we see that both the real and the imaginary part of $A_1$ are zigzag on arbitrarily minute scales. In fact, the imaginary part of $A_1$ is proportional to the \emph{Riemann function},
\begin{eqnarray}
   R(t) =\sum_{n=1}^\infty \frac{\sin n^2 t}{n^2}  .
\end{eqnarray}
Riemann conjectured that this function is nowhere differentiable. In 1916, Hardy proved that it is indeed not differentiable when $t$ is an irrational multiple of $\pi $ \cite{hardy}. But the problem was completely solved only much later. Around 1969, Gerver proved that $R(t)$ is differentiable when $t = p\pi/q$ with $p$ and $q$ being odd integers, and not differentiable elsewhere \cite{gerver, gerver2}. Hence, we know that $A_1$ is differentiable at most at countably many points.

\begin{figure}[tb]
\centering

\includegraphics[width= 0.45\textwidth ]{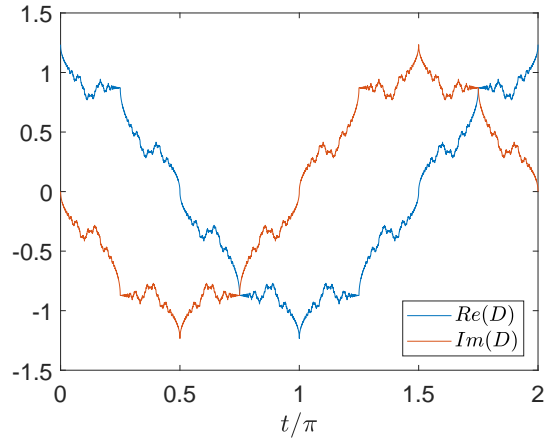}
\caption{(Color online) Real and imaginary parts of the function $D(t)$ in (\ref{D}). The first and second derivative of $A_1$ and $A_2$, respectively, are proportional to $D$. We have $Re(D) \sim \pi^2/8 - \sqrt{\pi t/8 } $ and $Im(D)\sim -\sqrt{\pi t/8 }$ asymptotically as $t\rightarrow 0$.}
\label{fig_D}
\end{figure}

As for $A_2$ and $A_3$, term-by-term differentiation is legitimate once and twice, respectively, as the resultant series, are proportional to
\begin{eqnarray}\label{D}
   D(t) =\sum_{n\geq 1, n\in {odd}}  \frac{e^{ -i  n^2 t}}{n^2} ,
\end{eqnarray}
which is uniformly convergent on $\mathbb{R}$. However, further differentiation seems impossible, as hinted by the plot of $D$ (see figure~\ref{fig_D}) and by its similarity with the Riemann function.

Hence, in accordance with the increasing smoothness of the initial state $\psi_i$ ($1\leq i \leq3$), the autocorrelation functions $A_i$ ($1\leq i \leq3$) are increasingly smooth. We have plotted them in  figure~\ref{fig2} for comparison. There, we see that the graphs of $A_1$ look spiky and fractal. The graphs of $A_2$ are no longer spiky, but the wiggles are apparent. Only for $A_3$, there is no visible nonsmoothness---Due to the relative fast decay of the coefficients in (\ref{autofunscc}), the first term dominates and therefore, the curves of $A_3$ look  sinusoidal. We note that Fourier series like those in (\ref{autofunsaa})-(\ref{autofunscc}) were studied by Chamizo and C\'ordoba decades ago \cite{fractal}. By their general theorem, the graphs of $A_1$ have a fractal (box-counting) dimension of $5/4$, while those of $A_2$ and $A_3$ have a dimension of unity.

\section{Short-time behavior}
Apart from the overall smoothness, there is yet another difference between the curves of $A_1$ and $A_3$ that is impossible to miss. In figure~\ref{fig2}(a), at $t=0$, the curves drop down almost vertically in the initial phase and develop some well-shaped cusps. In contrast, in figure~\ref{fig2}(c), the curves go either horizontally, or downwards but with a finite slope. This observation motivates us to study the asymptotic behaviors of the autocorrelation functions in the short time limit. Note that other than differentiability, asymptotics provides a different perspective and an independent way to characterize the smoothness or nonsmoothness of a function. A function like $A_1$ which is poor in differentiability could well approach a smooth function asymptotically.

\begin{figure}[tb]
\centering
\includegraphics[width= 0.45\textwidth ]{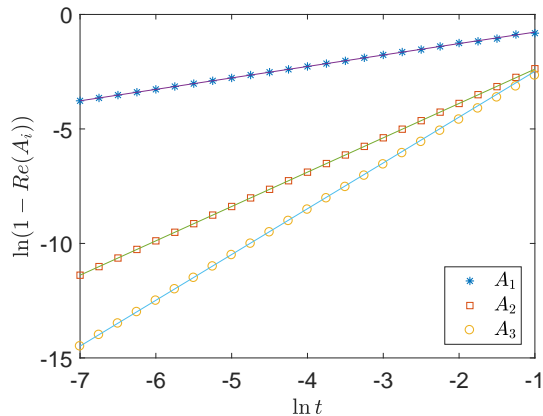}
\caption{(Color online) Checking the validity of the asymptotic formulas in (\ref{asymptotics1aa})-(\ref{asymptotics1ff}). The markers correspond to numerically exact values, while the straight lines indicate the leading term in each analytic formula. For the sake of clarity, only the real parts of the $A$'s are displayed here, but  the agreement between the numerical values and the analytic predictions is equally good for the imaginary parts.    }
\label{fig_Ascale}
\end{figure}

The calculation is deferred to \ref{appendixA}. There, by using the Mellin transform \cite{mellin, hughes}, we get that \cite{comment}
\begin{eqnarray}
  Re(A_1) &\sim & 1 - \sqrt{\frac{18}{\pi^3}} t^{1/2} + O(t^{3/2}), \;\;  \label{asymptotics1aa} \\
   Im(A_1) &\sim & - \sqrt{\frac{18}{\pi^3}} t^{1/2} + \frac{3}{\pi^2} t +O(t^{3/2}),\quad\; \label{asymptotics1bb}  \\
  Re(A_2) &\sim & 1 - \sqrt{\frac{512}{\pi^7}} t^{3/2} +O(t^3), \;\; \label{asymptotics1cc} \\
   Im(A_2) &\sim  &  - \frac{12}{\pi^2} t + \sqrt{\frac{512}{\pi^7}} t^{3/2}+O(t^3) , \quad \label{asymptotics1dd} \\
  Re(A_3) &\sim & 1 - \frac{60}{\pi^4}t^{2} + \sqrt{\frac{8192}{\pi^{11}}} t^{5/2} +O(t^3), \quad\quad\;  \label{asymptotics1ee}\\
    Im(A_3)& \sim &  - \frac{10}{\pi^2} t +  \sqrt{\frac{8192}{\pi^{11}}} t^{5/2} +O(t^3)  ,   \quad \label{asymptotics1ff}
\end{eqnarray}
as $t\rightarrow 0$. The correctness of these formulas can be and is confirmed by numerics, as done in figure \ref{fig_Ascale}.

The $t^{1/2}$ terms in $A_1$ account for the steepness of the curves at $t= 0$ in figure~\ref{fig2}(a).
Note that unlike $|A_3 |^2$, the survival probabilities $|A_1|^2$ and $|A_2|^2$ behave nonquadratically for short times. Such fractional power law behavior was also noticed in previous studies \cite{muga,  schuss, nonquadratic, pons}, and is of great interest in the context of quantum Zeno effect. As argued by Muga \emph{et al}. \cite{muga}, here the $t^{1/2}$ behavior of $A_1$ is linked to the fact that $\langle \psi_1 |H_{{w}} |\psi_1 \rangle $ diverges, and the $t^{3/2}$ behavior of $A_2$  to the fact that $\langle \psi_2 |H_{{w}}^2 |\psi_2 \rangle $ diverges.
Indeed, from the point of Fourier analysis, for a generic hamiltonian $H$ and a generic initial state $|\psi_i \rangle $ in (\ref{defa}), the existence of the $n$th momentum of the hamiltonian $\langle \psi_i |H^n | \psi_i \rangle $ means that the Fourier series in the definition (\ref{forma}) of the autocorrelation function  can be differentiated term-by-term for $n$ times, and the resulting $n $th derivative $A^{(n)} (t)$ is a continuous and bounded function of $t$. That is, $A(t)$, and hence the survival probability $|A(t)|^2$, are of class $C^n $. This rules out a $t^{\mu}$ short-time behavior of $|A|^2$ for $\mu < n $.

In our specific case, apparently,  $\langle \psi_1 |H_{{w}} |\psi_1 \rangle $ and $\langle \psi_2 |H_{{w}}^2 |\psi_2 \rangle $ diverge because in the series (\ref{psiexpanaa}) and (\ref{psiexpanbb}), the coefficients decay too slowly. But this in turn is because $\psi_1$  and $\psi_2'$ are discontinuous. Hence, we see that  the smoothness or nonsmoothness of an initial state affects not only the differentability of the autocorrelation function as a whole but also its short-time behavior.

This point can be best illustrated with a generalization of the state $\psi_1 $. Let us consider the state depending on a parameter $0<\beta < 1/2$ \cite{comment2},
\begin{eqnarray}\label{chibeta}
  \chi_\beta &=& \frac{1}{\sqrt{\zeta(2+ 2 \beta )}}\sum_{n\geq 1} \frac{\phi_n}{n^{1+\beta}} ,
\end{eqnarray}
where in the normalization factor we have the Riemann zeta function. Its autocorrelation function is
\begin{equation}\label{autofunchi}
  A_\beta (t ) = \frac{1}{\zeta(2+ 2 \beta )}\sum_{n\geq 1} \frac{e^{-i n^2 t}}{n^{2+2\beta}} .
\end{equation}
Now the single parameter $\beta $ controls the smoothness of the function $\chi_\beta$, the differentiability of the autocorrelation function $A_\beta$, as well as its short-time behavior.
Again by using the Mellin transform, we have (see \ref{appendixB})
\begin{equation}\label{smallx}
  \chi_{\beta} \sim \frac{\Gamma(-\beta) \sin(-\pi \beta /2 ) \sqrt{2} }{\sqrt{ \pi \zeta(2+ 2 \beta )}} x^\beta
\end{equation}
as $x\rightarrow 0 $. Hence, the discontinuity of $\psi_1 $ at $x =0$ is softened into a power law behavior $x^\beta$. The nonsmoothness or singularity of the function $\chi_\beta $ is revealed in its first derivative, which diverges like $x^{\beta -1}$ as $x\rightarrow 0 $. The smaller $\beta$, the stronger the singularity. As for the differentiability of $A_\beta$, according to Chamizo and C\'ordoba \cite{fractal}, for $\beta < \beta_c = 1/4$, the graphs of $A_\beta$ have a fractal dimension larger than unity, which means that it is nondifferentiable almost everywhere, while for $\beta > \beta_c = 1/4$, $A_\beta $ is absolutely continuous and hence differentiable almost everywhere. For the short-time behavior, we have that
 \begin{eqnarray}
   1- Re(A_\beta)  &\sim  & -\frac{ \Gamma(-\lambda)\cos (\lambda\pi /2) }{2\zeta(2+ 2\beta)} t^{\lambda},  \label{asymptoticschiaa} \\
   Im(A_\beta) &\sim & \frac{ \Gamma(-\lambda )\sin(\lambda \pi /2) }{2\zeta(2+ 2\beta)} t^{\lambda} , \label{asymptoticschibb}
 \end{eqnarray}
as $t\rightarrow 0 $. Here the point is that the exponent $\lambda = (1+ 2 \beta )/2$ can be tuned continuously between $1/2$ and $1$, while previously only the values $1/2$ and $3/2$ were observed \cite{muga, pons, nonquadratic, schuss}.

\begin{figure}[tb]
\centering
\includegraphics[width= 0.45\textwidth ]{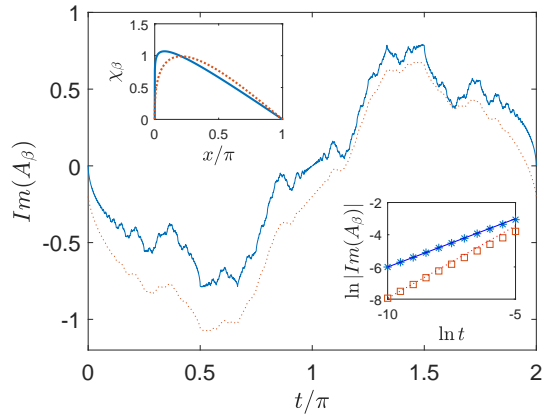}
\caption{(Color online) Imaginary part of the autocorrelation function $A_\beta $ of the state $\chi_\beta$ [see (\ref{chibeta})-(\ref{asymptoticschibb})]. The upper inset illustrates the two wave functions $\chi_\beta$, while the lower inset demonstrates the $t^{(1+2\beta )/2}$ short-time behavior of $A_\beta$ (the straight lines are of slope $(1+2\beta )/2$). The solid lines correspond to the value of $\beta = 0.1$, while the dotted lines the value of $\beta = 0.4 $. For clarity, in the main frame, the dotted line is shifted downwards by $0.2$.  }
\label{fig_Abeta}
\end{figure}

All these are verified in figure~\ref{fig_Abeta}, where we have examined two cases with $\beta = 0.1 <\beta_c $ and $\beta = 0.4 > \beta_c$. In the upper inset, we see that as $\beta $ increases, the singularity of the wave function at $x = 0$ softens and the boundary layer gets thicker. Correspondingly, the curve of the autocorrelation functions gets smoother---The curve with $\beta = 0.1$ is visibly fractal; the one with $\beta = 0.4 $ is still not very smooth but apparently  no longer fractal. As for the short-time behavior, as the lower inset shows, both curves follow the predicted power law, with the exponent increasing with the parameter $\beta $ linearly. Hence, we see that by tuning the singularity of the initial state, i.e., by changing the value of $\beta $, we can tune both the differentiability and short-time behavior of the autocorrelation function.

\begin{figure*}[tb]
\centering
\includegraphics[width= 0.45\textwidth ]{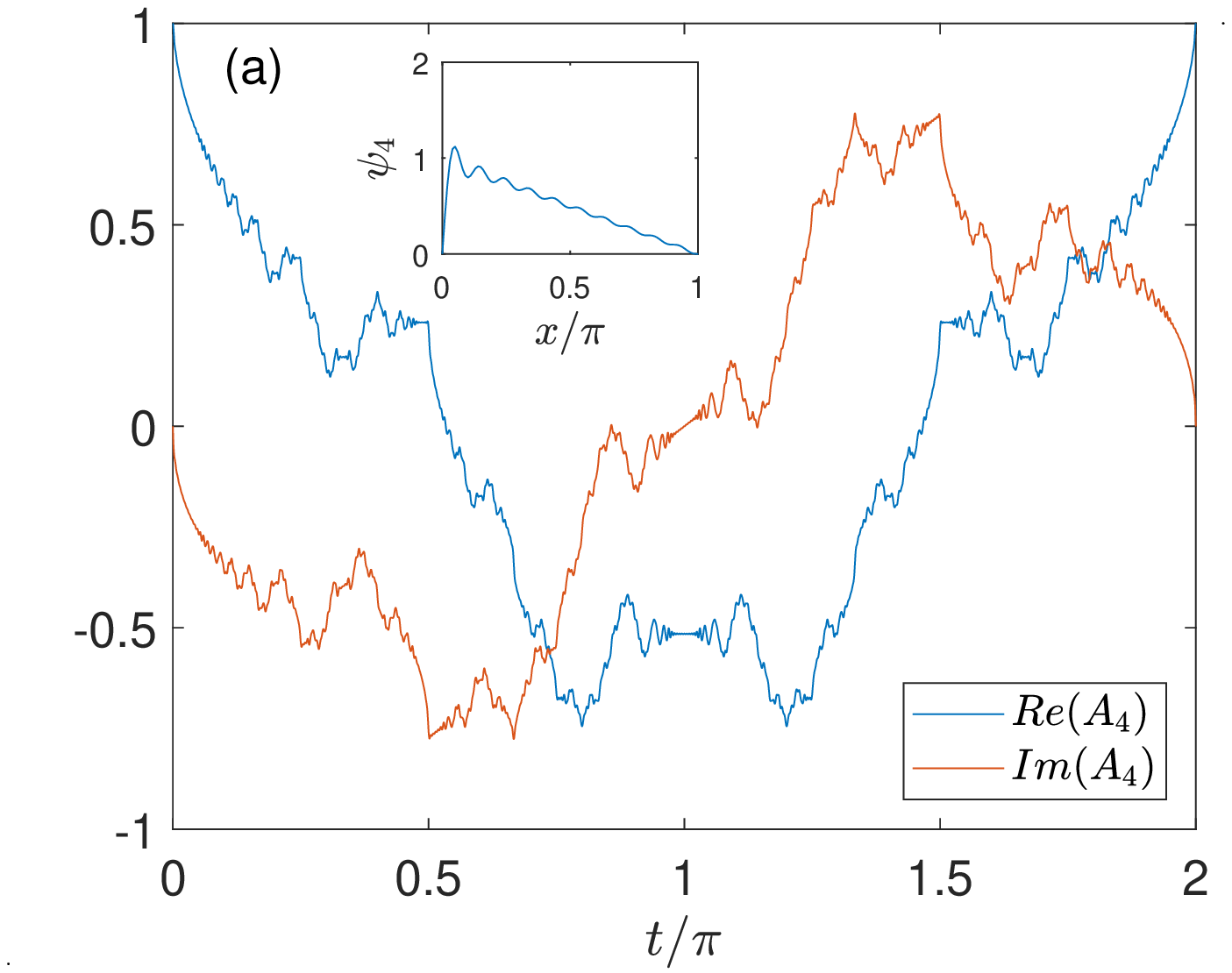}
\includegraphics[width= 0.45\textwidth ]{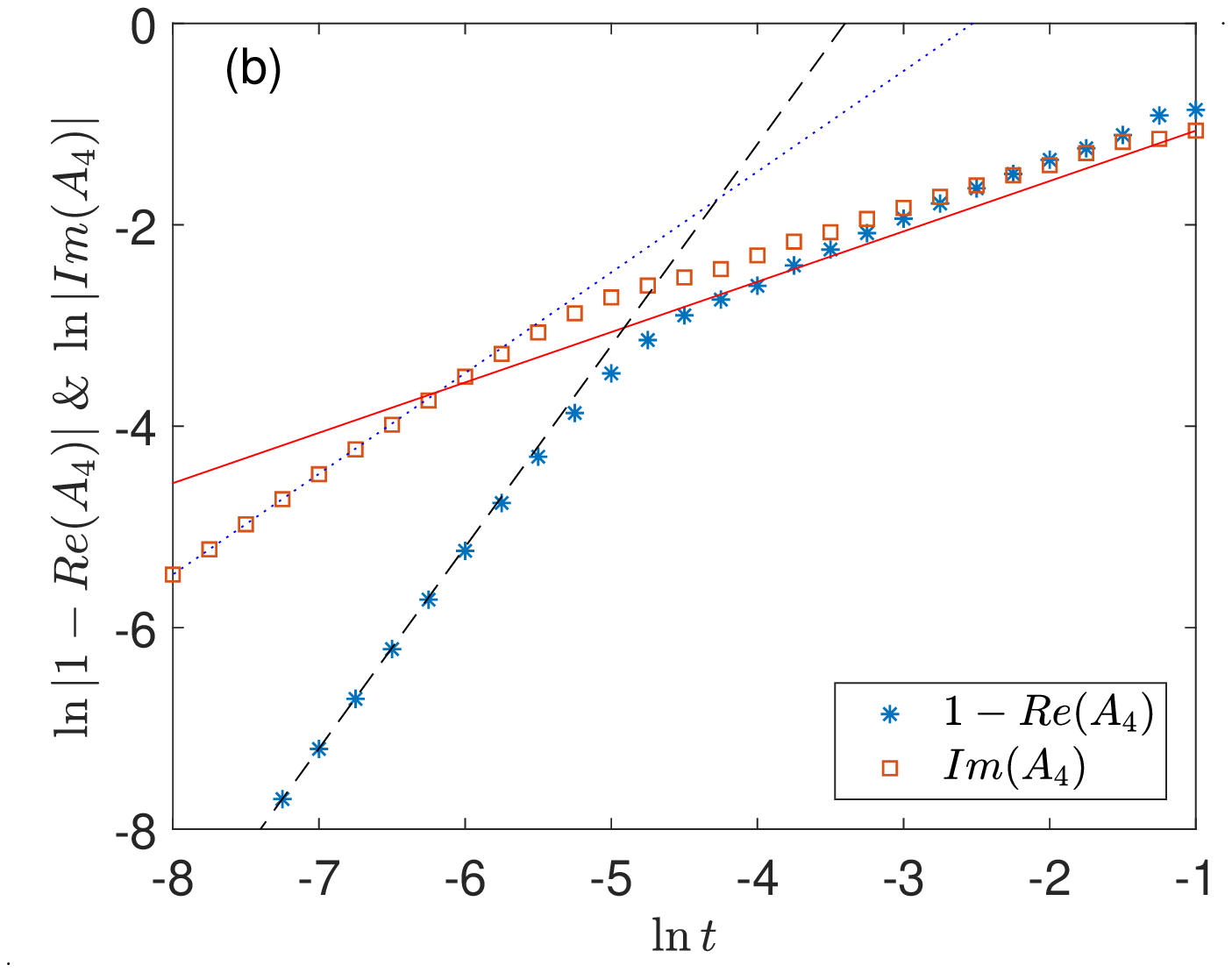}
\caption{(Color online) (a) Real and imaginary parts of the autocorrelation function of the state $\psi_4$ in (\ref{psi4}), which is illustrated in the inset. Compare the curves with those in figure~\ref{fig2}(a). (b) For $t$ neither too small nor too large, the autocorrelation function follows the $t^{1/2} $ scaling law in (\ref{asymptotics1aa}) and (\ref{asymptotics1bb}) approximately. The solid line is of slope $1/2$, the dotted line is of slope $1$, while the dashed line is of slope 2.  }
\label{fig_A4}
\end{figure*}

\section{Problem of relevance}
Finally, let us address some issue with the states $\psi_{1}$ and $\psi_{2}$ (and $\chi_\beta$ as well). An objection might be that these states are not legitimate states of the infinite square potential.
In \cite{gitman}, the domain $\mathcal{D}$ of the infinite square well hamiltonian is defined in a rigorous way. An important point is that, for a state $\psi$ to belong to the domain $\mathcal{D}$, the wave function $\psi $ itself and its first derivative $\psi' $ should be absolutely continuous. But here $\psi_1$ and $\psi_2' $ are discontinuous somewhere, and hence not absolutely continuous. Actually, when some hamiltonian $H $ acts on a state $\psi_i  $ belonging to its domain, the product $H\psi_i  $, although not necessarily still in the domain, at least should be in the Hilbert space. That is, the series
\begin{eqnarray}
  \| H \psi_i \|^2 &=& \sum_{n=1}^\infty  |c_n|^2 E_n^2
\end{eqnarray}
should converge. But this is exactly the second moment of $H $ with respect to $\psi_i $, $\langle \psi_i |H^2 | \psi_i \rangle $. Therefore, by our arguments above, \emph{a legitimate state cannot decay nonquadratically asymptotically.}

All these are reasonable. But the point is that, the mere fact that a state does not belong to the domain of a hamiltonian does not render the studying of its dynamics meaningless. An invalid state can be approximated to arbitrary precision by valid states, and accordingly, the peculiarities of its dynamics can be retained by these states. For example, by truncating the series in (\ref{psiexpanaa}), we can get a totally valid state
\begin{eqnarray}\label{psi4}
  \psi_4 &\propto &  \sum_{n=1}^{20} \frac{\sqrt{6} }{n \pi} \phi_n .
\end{eqnarray}
This state, as shown in the inset of figure~\ref{fig_A4}, resembles the state $\psi_1$,  but avoids its difficulty by satisfying the boundary condition and being infinitely smooth. As a neighbor of $\psi_1$ in the Hilbert space, its autocorrelation function $A_4(t)$, which is a truncation of the $A_1$ series in (\ref{autofunsaa}), retains the fractal features of the latter to a very small time scale. Moreover, although in an initial phase when $t$ is sufficiently small, $|A_4|^2$ decays quadratically, in a subsequent interval with significant duration, it follows the $t^{1/2}$ feature of $|A_1|^2$ well. This is visible both in figure \ref{fig_A4}(a) and figure \ref{fig_A4}(b).

\section{Conclusions and discussions}

In conclusion, by borrowing mathematical results from Fourier analysis, we have gained some insights into quantum dynamics.
The first message is that less smooth wave functions (i.e., wave functions with jumps or cusps or other singularities) lead to autocorrelation functions with poor differentiability. In some extremal cases, the autocorrelation function can even be nondifferentiable almost everywhere.  However, the nonsmoothness or singularity of the initial state affects not only the overall differentiability of the autocorrelation function, but also its short-time behavior, which is a primary concern in the discussion of quantum Zeno effect. Specifically, nonsmooth or singular wave functions could lead to fractional power law behavior of the autocorrelation function in the short-time regime, resulting in deviation from the quadratic decay assumption. Indeed, we notice that in all previous reports of nonquadratic decay \cite{muga, schuss, nonquadratic,  pons}, the initial states are either with jumps or with cusps. The new insight we have gained here is that the exponent of the fractional power law can be tuned continuously by tuning the singularity of the initial state, while previously only the two discrete values of $1/2$ and $3/2$ were known. Moreover, now we know that in the asymptotic expansion of the autocorrelation function,  other than the powers $t^a $, we might have terms like $t^a \ln t $ too.

Some remarks are in order.

First, in our discussion, we have tacitly assumed that all the eigenstates are bound states. But it is conceivable that similar connections exist when some or all of the eigenstates are extended. 

Second, in the infinite square well case, it is the Fourier theory in the narrow sense that takes a part. For other systems, for example, the harmonic oscillator, the eigenstates are different, and the Fourier expansion is in a board sense. But as a general fact, the ``smoothness means fast decay'' principle still holds \cite{davis,wang,xiang}.

Third, in the story, of paramount importance is the decay mode of the coefficients $|c_n|^2$. Although in our cases, their calculation is easy, it is conceivable that in other problems, determining their values or their decay order could be a technical challenge.

Fourth, in our tentative work, we have confined ourself to the one dimensional world. Generalization to higher dimensions seems both interesting and challenging. Many difficulties would pop up, one of which is to deal with multiple Fourier series.

Fifth, while here we have taken some toy model to illustrate our point, the idea is definitely relevant in more realistic problems. For example, in solid state physics, we have the phenomenon of Friedel oscillations \cite{friedel}---In a Fermi sea, the density ripples around an impurity decay slowly in a power law. This characteristics can be easily understood or even anticipated from our perspective. The density distribution in the real space is essentially the Fourier transform of the distribution function in the momentum space, which has a jump at the fermi level. As yet another example, in Fourier optics, we know that the Fraunhofer diffraction patter of a slit is the sinc function (its square actually) \cite{fraunhofer}, which decays as $1/x$. This slow decay is because the window function has jumps at the two ends.

Sixth, to our delight, now we understand the periodic \emph{cusps} of the autocorrelation function in the quench dynamics of a Bloch state  \cite{epl1,epl2}, which was observed and even quantitatively accounted for but not really understood. In that problem, the autocorrelation function $A_5(t) = e^{-i\alpha t } B(t)$, with $\alpha $ some non-integral real number and the $2\pi$-periodic function $B(t)$ defined by the series \cite{ejp},
\begin{eqnarray}
 B(t) =  \frac{\sin^2 \pi \alpha}{\pi^2} \sum_{n\in \mathbb{Z}} \frac{e^{-in t }}{(n+\alpha )^2} .
\end{eqnarray}
Here the weight coefficients $|c_n|^2 = \sin^2 (\pi \alpha)/[\pi^2 (n+\alpha)^2 ] $ decay like $1/n^2$. Because of this slow decay rate, we are to anticipate that the  function  $B(t)$ has some discontinuities in its derivatives. Indeed, it has the closed expression
\begin{eqnarray}
 B(t) =  \left( 1- \frac{1- e^{-i 2\pi \alpha }}{2\pi}t \right)e^{i\alpha t } , \quad 0\leq t \leq 2 \pi ,
\end{eqnarray}
and shows cusps whenever $t$ is an integral multiple of $ 2 \pi$. The survival probability $|A_5|^2$ decreases linearly (thus nonquadratically) in the short-time regime.

Finally, we should note that the link between autocorrelation function and Fourier analysis was noted decades ago by Krylov and Fock \cite{fock}. They realized that the autcorrelation function is simply the Fourier transform of the so-called energy distribution function. Later, Khalfin applied the Paley-Wiener theorem in Fourier analysis \cite{stein2}, which  exemplifies the principle, to this transform, and came to the important conclusion that exponential decay at all time is impossible for a realistic system with a spectrum bounded from below \cite{khalfin}. As the most natural way to view the autocorrelation function is seeing it as a Fourier series, we believe that the field of quantum dynamics will continually benefit from the theory of Fourier analysis, which is full of profound results.

\section*{Acknowledgments}
We are grateful to J. Guo, K. Yang, R. Huang,  G. Yang, and Y. D. Hu for helpful discussions. This work is supported by the National Science Foundation of China under Grant No. 11704070.

\appendix

\section{Derivation of the short-time asymptotics in (\ref{asymptotics1aa})-(\ref{asymptotics1ff})}{\label{appendixA}}

The series  defining $A_{1\leq i \leq 3} $ in (\ref{autofunsaa})-(\ref{autofunscc}) and $D$ in (\ref{D}) are somehow similar, and can be treated in essentially  the same way. To do so, we shall consider some more general problems, i.e., the functions
\begin{eqnarray}
  f_1(x;\mu) &=& \sum_{n \geq 1 } \frac{e^{-i n^2 x^2 }}{n^\mu}, \\
   f_2(x;\mu) &=& \sum_{n \geq 1 , n\in {odd}} \frac{e^{-i n^2 x^2 }}{n^\mu} ,
\end{eqnarray}
where $ \mu > 1 $ is a parameter characterizing the decay rate of the coefficients. Note that the $A$'s and the $D$ are either proportional to $f_1$ or to $f_2$ if we set $x = t^{1/2}$ and choose a proper value of $\mu $.

In the definitions of $f_{1,2}$, we have the so-called harmonic summation \cite{mellin}. For problems in this category, the Mellin transform is useful.  Let $h(x)$ be a function defined on $(0,\infty )$. Its Mellin transform $\mathcal{M}[h]$ is a function of the complex variable $s = \sigma + i \eta $ defined by the integral,
\begin{eqnarray}\label{transform1}
  \mathcal{M}[h; s ] &=& \int_0^\infty x^{s-1} g(x) dx .
\end{eqnarray}
The integral converges in some vertical strip $\alpha < Re(s) < \beta $. This is then the domain of $\mathcal{M}[h]$. The inversion formula is
\begin{eqnarray}\label{inversion}
  h(x) &=& \frac{1}{2\pi i } \int_{\mathcal{C}_\sigma} x^{-s} \mathcal{M}[h;s] d s  ,
\end{eqnarray}
where $\mathcal{C}_\sigma$ is the vertical line from $\sigma -i \infty $ to $\sigma + i \infty $, with $\alpha < \sigma < \beta$.

Let us focus on $f_1$ first, and treat the real and imaginary parts independently. By the linearity of the transform (\ref{transform1}), we have readily that
\begin{eqnarray}
  \mathcal{M}[Re(f_1); s ] &=& \zeta(s+ \mu ) \mathcal{M}[\cos x^2; s ] , \quad \label{zeta1aa} \\
  -\mathcal{M}[Im(f_1); s ] &=& \zeta(s+ \mu ) \mathcal{M}[\sin x^2; s ]  .\quad \label{zeta1bb}
\end{eqnarray}
where $\zeta $ is the Riemann zeta function. Using the known result that ($ 0<Re(s) < 1$) \cite{stein2}
\begin{eqnarray}
  \mathcal{M}[\cos x; s ]&=& \Gamma(s) \cos \frac{\pi s}{2} ,\\
  \mathcal{M}[\sin x; s ]&=& \Gamma(s) \sin \frac{\pi s}{2} ,
\end{eqnarray}
by change of variables, we get easily ($ 0<Re(s) < 2$)
\begin{eqnarray}
  \mathcal{M}[\cos x^2; s ]&=& \frac{1}{2}\Gamma\left(\frac{s}{2} \right) \cos \frac{\pi s}{4} ,   \\
  \mathcal{M}[\sin x^2; s ]&=& \frac{1}{2} \Gamma\left (\frac{s}{2} \right) \sin \frac{\pi s}{4} .
\end{eqnarray}
Substituting these into (\ref{zeta1aa})-(\ref{zeta1bb}) and then invoking the inversion formula (\ref{inversion}), we get finally the integral representations of $f_1$ as
\begin{eqnarray}
  Re(f_1(x)) = \frac{1}{4\pi i }  \int_{\mathcal{C}_1} x^{-s} \Gamma\left(\frac{s}{2} \right) \cos \frac{\pi s}{4}  \zeta(s+ \mu ) ds,\quad \quad  \label{f1expaa} \\
  Im(f_1(x)) =\frac{-1}{4\pi i }  \int_{\mathcal{C}_1} x^{-s} \Gamma\left(\frac{s}{2} \right) \sin \frac{\pi s}{4}  \zeta(s+ \mu ) ds.  \quad\quad \label{f1expbb}
\end{eqnarray}
Although initially the integrands are defined only in the strip $ 0<Re(s) < 2$, they extend analytically into the whole complex plane as meromorphic functions. By the asymptotic behaviors of the integrands, we can shift the integration line to the left. In this process, the poles of the integrands are rounded up one by one, and we get an asymptotic expansion of the original function for small $x$.

To be specific, let us consider $f_1$ with $\mu = 2$, which is proportional to $A_1$. In (\ref{f1expaa}), from right to left, there are two poles, $s=0$ and $s= -1$, in which  the pole $s = 0 $ is due to the gamma function, while the pole $s =-1$ is due to the zeta function \cite{cancel}. By shifting the integration line from $\mathcal{C}_1 $ to $\mathcal{C}_{-2}$, which is justified by the asymptotics of the gamma function, the cosine function, and the zeta function as $|\eta | \rightarrow \infty $,  we get
\begin{eqnarray}
  Re(f_1(x ; 2)) &\sim & \zeta(2) + \frac{1}{2\sqrt{2}} \Gamma\left(-\frac{1}{2} \right)  x + O(x^2) \nonumber \\
  & =& \frac{\pi^2}{6} - \sqrt{\frac{\pi}{2}} x +O(x^2 ) , \;\;\;  x \rightarrow 0. \quad  \quad
\end{eqnarray}
In (\ref{f1expbb}), from right to left, there are two poles, $s=-1$ and $s= -2$, in which the pole $s =-1$ is due to the zeta function, while the pole $s = -2 $ is due to the gamma function. By shifting the integration line from $\mathcal{C}_1 $ to $\mathcal{C}_{-3}$, we get
\begin{eqnarray}
  Im (f_1(x ; 2)) &\sim & \frac{1}{2\sqrt{2}} \Gamma\left(-\frac{1}{2} \right)  x - \zeta(0 )x^2 +O(x^3) \nonumber \\
  &=& -  \sqrt{\frac{\pi}{2}} x +\frac{1}{2}x^2 +O(x^3 )  , \;\;\;  x \rightarrow 0.\quad
\end{eqnarray}
Taking into account the normalization factor $1/\zeta(2) = 6/\pi^2 $ and the fact that $x = t^{1/2 }$, we obtain (\ref{asymptotics1aa}) and (\ref{asymptotics1bb}).

The treatment of $f_2$ is similar. The only difference is that restricting the summation to odd numbers introduces a factor of $(1- 2^{-s- \mu })$. That is, instead of (\ref{zeta1aa})-(\ref{zeta1bb}), we have
\begin{eqnarray}\label{zeta2}
  \mathcal{M}[Re(f_2); s ] &=&  (1- 2^{-s- \mu })\zeta(s+ \mu ) \mathcal{M}[\cos x^2; s ] , \quad\quad\quad \\
  -\mathcal{M}[Im(f_2); s ] &= &(1- 2^{-s-\mu }) \zeta(s+ \mu ) \mathcal{M}[\sin x^2; s ]  .\quad\quad
\end{eqnarray}
By the same procedure of enumerating the poles and calculating the residues, we get (\ref{asymptotics1cc})-(\ref{asymptotics1ff}).

From the calculation above, we see that for a generic $\mu$, the pole of the zeta function will results in a  $t^{(\mu-1)/2}$-term in both the real and imaginary parts of the autocorrelation function. In the real part, only for $\mu> 5$, is this term dominated by the $t^2$-term; in the imaginary part, only for $\mu>3$, is it dominated by the $t$-term. Therefore, only for $\mu>5$, as in the case of $\psi_3 $, we have a quadratic short-time behavior of the survival probability. This is also the critical value of $\mu $ when the second moment of the hamiltonian converges.
For smaller $\mu $, as in the case of $\psi_{1,2}$, we get nonquadratic decay.

But the main point here is that the short-time power law behavior of the autocorrelation function can be tuned by adjusting the value of $\mu $. Exponents other than the values of $1/2$ and $3/2$ can be realized too.

Another point is that when two simple poles merge into a double pole, we might get terms like $x^a (\ln x)^b $ in the asymptotic expansions.  For example, if $\mu = 3$, $s = -2$ is a double pole of $\mathcal{M}( Im (f_1) )$. The leading term of the asymptotic expansion of $Im(f_1)$ is then proportional to $x^2 \ln x $, or $t \ln t $.

\section{Asymptotics of the wave function $\chi_\beta$}{\label{appendixB}}

The asymptotics of $\chi_\beta $ in (\ref{chibeta}) can be treated similarly. Essentially, we have to consider the function defined as
\begin{eqnarray}
  f_3(x;\mu) &=& \sum_{n \geq 1 } \frac{\sin n x }{n^\mu} ,\quad \mu =1 + \beta .
\end{eqnarray}
Its Mellin transform is
\begin{eqnarray}
  \mathcal{M}[f_3; s ]&=& \Gamma(s) \sin \frac{\pi s}{2} \zeta(s + 1+ \beta ) , \quad 0<Re(s) < 1.
\end{eqnarray}
The function $f_3$ is recovered as
\begin{eqnarray}
  f_3(x)=\frac{1}{2\pi i }  \int_{\mathcal{C}_{1/2}} x^{-s} \Gamma(s) \sin \frac{\pi s}{2} \zeta(s + 1+ \beta ) ds.  \quad\quad
\end{eqnarray}
The zeta function contributes a pole $s = - \beta $. The factor $\Gamma(s) \sin \frac{\pi s}{2} $ has poles at $s = -1 , -3, -5$, etc. Hence, for $\beta \in (0, 1)$, the leading term in the asymptotic expansion of $\chi_\beta $ is proportional to $x^\beta $.


\section*{References}

\begin{thebibliography}{99}

\bibitem{robinett1}
  Robinett R W
  2004
  Quantum wave packet revivals
  {\it  Phys. Rep.}
  {\bf 392}
  1

\bibitem{nauenberg}
  Nauenberg M
  1990
  Autocorrelation function and quantum recurrence of wavepackets
  {\it J. Phys. B: At. Opt. Phys.}
  {\bf 23}
  L385

\bibitem{stein}
  Stein E M and Shakarchi R
  2003
  {\it Fourier Analysis}
  (Princeton University Press)

\bibitem{hardybook}
  Hardy G H and Rogosinski W W
  2013
  {\it Fourier Series}
  (Dover, New York)

\bibitem{bounded}
  Taibleson M
  1967
  Fourier coefficients of functions of bounded variation
  {\it Pro. Amer. Math. Soc.}
  {\bf 18}
  766

\bibitem{zero}
  Gilbert L P, Belloni M, Doncheski M A and Robinett R W
  2006
  Playing quantum physics jeopardy with zero-energy eigenstates
  {\it Am. J. Phys.}
  {\bf 74}
  1035

\bibitem{knopp}
  Knopp K
  1990
  {\it Theory and Application of Infinite Series}
  (Dover, New York)

\bibitem{hardy}
  Hardy G H
  1916
  Weierstrass's non-differentiable function
  {\it Trans. Am. Math. Soc.}
  {\bf 17}
  322

\bibitem{gerver}
  Gerver J
  1970
  The differentiability of the Riemann function at certain rational multiples of $\pi$
  {\it Am. J. Math.}
  {\bf 92}
  33

\bibitem{gerver2}
  Gerver J
  1971
  More on the differentiability of the Riemann function
  {\it Am. J. Math.}
  {\bf 93}
  33

\bibitem{fractal}
  Chamizo F and Cordoba A
  1999
  Differentiability and dimension of some fractal Fourier series
  {\it Adv. Math.}
  {\bf 142}
  335

\bibitem{mellin}
  Flajolet P, Gourdon X and Dumas P
  1995
  Mellin transforms and asymptotics: Harmonic sums
  {\it Theoretical Computer Science}
  {\bf 144}
  3

\bibitem{hughes}
  Ninham B W, Hughes B D, Frankel N E and Glasser M L
  1992
  M{\"o}bius, Mellin, and mathematical physics
  {\it Physica A}
  {\bf 186}
  441

\bibitem{comment}
  {The fractal exponent $1/2$ of the leading term of (\ref{asymptotics1bb})
is the best indicator that the corresponding series (\ref{autofunsaa}) decays slowly. Any
finite truncation of the series should have an asymptotic expansion (a Taylor expansion actually)
starting with a term proportional to $t$ in the imaginary part. That this is not the case for the
infinite series signals that as $t\rightarrow 0$, the tail of the series dominates.}

\bibitem{muga}
  Muga J G, Wei G W and Snider R F
  1996
  Short-time behaviour of the quantum survival probability
  {\it EPL}
  {\bf 35}
  247

\bibitem{schuss}
  Marchewka A and Schuss Z
  2000
  Path-integral approach to the Schr{\"o}dinger current
  {\it Phys. Rev. A}
  {\bf 61}
  052107

\bibitem{nonquadratic}
  Cordero S and Garc{\'\i}a--Calder{\'o}n G
  2012
  Analytical study of quadratic and nonquadratic short-time behavior of quantum decay
  {\it Phys. Rev. A}
  {\bf 86}
  062116

\bibitem{pons}
  Sokolovski D, Pons M and Kamalov T
  2012
  Anomalous Zeno effect for sharply localized atomic states
  {\it Phys. Rev. A}
  {\bf 86}
  022110

\bibitem{comment2}
  {Here we restrict $\beta $ to $\beta <1/2$ simply for the ease
of narrative. For $\beta > 1/2$, the leading term in the asymptotic expansion
of $Im(A_\beta )$ is proportional to $t$, instead of being proprotional to $t^{(1+2 \beta)/2}$ as
in (\ref{asymptoticschibb}). The reason is that as $\beta $ crosses $1/2$, two simple
poles ($s = -2$ and $s =-1 - 2 \beta $) of the Mellin transform of $Im(A_\beta)$ changes order.
At $\beta = 1/2$, they coincide and merge into a double pole and leading term in the asymptotic
expansion of $Im(A_\beta )$ is proportional to $t \ln t $}

\bibitem{gitman}
  Gitman D M, Tyutin I V and Voronov B L
  2012
  {\it Self-Adjoint Extensions in Quantum Mechanics}
  (Springer, Berlin)

\bibitem{davis}
  Davis P J
  1975
  {\it Interpolation and Approximation}
  (Dover, New York)

\bibitem{wang}
  Wang H and Xiang S
  2012
  On the convergence rates of Legendre approximation
  {\it Math. Comp.}
  {\bf 81}
  861

\bibitem{xiang}
  Xiang S
  2012
 Asymptotics on Laguerre or Hermite polynomial expansions and their applications in Gauss quadrature
  {\it J. Math. Anal. Appl.}
  {\bf 393}
  434

\bibitem{friedel}
Friedel J 1952 The distribution of electrons round impurities in monovalent metals {\it Philos. Mag.} \textbf{43} 153

\bibitem{fraunhofer}
Goodman J 2005 {\it Introduction to Fourier Optics} 3rd ed. (Roberts \& Company Publishers)

\bibitem{epl1}
  Zhang J M and Yang H T
  2016
  Cusps in the quench dynamics of a Bloch state
  {\it EPL}
  {\bf 114}
  60001


\bibitem{epl2}
  Zhang J M and Yang H T
  2016
  Sudden jumps and plateaus in the quench dynamics of a Bloch state
  {\it EPL}
  {\bf 116}
  10008

\bibitem{ejp}
  Yang K L and Zhang J M
  2019
  On an exactly solvable toy model and its dynamics
  {\it Eur. J. Phys.}
  {\bf 40}
  035401

\bibitem{fock}
  Krylov N S and Fock V A
  1947
 The two interpretations of the uncertainty principle for energy and time
  {\it JETP}
  {\bf 17}
  93
  


\bibitem{stein2}
  Stein E M and Shakarchi R
  2003
  {\it Complex Analysis}
  (Princeton University Press)

\bibitem{khalfin}
  Khalfin L A
  1957
  Contribution to the decay theory of a quasi-stationary state
  {\it JETP}
  {\bf 33}
  1371

\bibitem{cancel}
  {Most poles of the gamma function are cancelled either by
  the zeros of the cosine/sine function or by those of the zeta function.}

\end{thebibliography}
\end{document}